\documentclass[twocolumn,prl,aps,showpacs,footinbib,amsmath,amssymb,superscriptaddress]{revtex4-1}
\usepackage{graphicx}
\usepackage{dcolumn}
\usepackage{xcolor}
\usepackage{bm}
\usepackage{natbib,hyperref}
\usepackage{amsmath,amssymb,amsfonts,bbold}
\usepackage[normalem]{ulem}
\newcommand{\cris}[1] {{\color{teal}{#1}}}

\begin{document}


\title{Moir\'e dispersion of edge states in spin chains on superconductors }

\author{Cristina Mier}
\affiliation{Centro de F{\'{\i}}sica de Materiales
        CFM/MPC (CSIC-UPV/EHU),  20018 Donostia-San Sebasti\'an, Spain}
\author{Deung-Jang Choi}
\affiliation{Centro de F{\'{\i}}sica de Materiales
        CFM/MPC (CSIC-UPV/EHU),  20018 Donostia-San Sebasti\'an, Spain}
\affiliation{Donostia International Physics Center (DIPC),  20018 Donostia-San Sebasti\'an, Spain}
\affiliation{Ikerbasque, Basque Foundation for Science, 48013 Bilbao, Spain}
\author{Nicol{\'a}s Lorente}
\email{nicolas.lorente@ehu.eus}
\affiliation{Centro de F{\'{\i}}sica de Materiales
        CFM/MPC (CSIC-UPV/EHU),  20018 Donostia-San Sebasti\'an, Spain}
\affiliation{Donostia International Physics Center (DIPC),  20018 Donostia-San Sebasti\'an, Spain}

\begin{abstract}
Our calculations of ferromagnetic spin chains on s-wave superconductors show that the energy oscillations of edge states with the chain’s length are due to a moir\'e pattern emerging from Friedel-like oscillations and  the discreteness of the spin-chain lattice. By modifying the spin lattice, the moir\'e dispersion of edge states can be controlled. In particular, we can engineer non-dispersive edge states that remain at fixed energy regardless of the size distribution of the spin chains. This is an important step in the study of edge states of spin chains that can be fabricated with a certain size dispersion.
\end{abstract}

\date{\today}

\maketitle
Majorana bound states (MBS) have gained increasing interest in recent years due to their non-Abelian statistics and potential application to topological quantum computing~\cite{Simon2008,Field_2018}.
The search for MBS in condensed matter systems pays particular attention to 1D structures that combine superconductivity, spin-orbit coupling (SOC) and magnetic interactions~\cite{Kitaev_2001,Hasan,Qi,Sato_review,Aguado_review,Choi_2019}. Under the right conditions, these structures are able to mimic a Kitaev chain where MBS  arise at the edges of finite systems~\cite{Kitaev_2001}.  Studies have focused on two main approaches: spin chains on superconducting surfaces~\cite{Yazdani, Yazdani2013, Ruby_2015,Pawlak_2016,Yazdani2017,Kim_2018,Choi_2019,Franke_arxiv,Schneider1, Schneider2}, and semiconducting nanowires with Rashba spin-orbit coupling and proximitized superconductivity~\cite{Hadas_2012,Rokhinson2012,Frolov_science, Churchil}. These edge states are expected to appear at zero-energy, however, the spatial overlap between MBS at both edges of finite systems results in non-zero-energy edge states~\cite{das_sarma_2009,das_sarma_2012,das_zero-bias_2012,klino_2012, Schneider2}. 

Due to the finite size of these systems, not only do MBS interact with each other but also with non-zero-energy states~\cite{Theiler_2019}. These interactions result on the mixing of MBS with the resulting states appearing at finite energies that quasi-periodically evolve with the system's size~\cite{Schneider1,Schneider2,Mier_topo}. Edge-state energy oscillations have also been found as a function of applied magnetic field~\cite{Churchil, Prada_2012,Cayao_2017}. Despite the mixing of states at the origin of the found oscillations, the new states maintain interesting topological properties~\cite{Leumer_2020}. Unfortunately, the oscillations in energy make the control of MBS challenging, and it would be desirable to find them at a fixed energy. Particularly, in large-scale manufacturing of spin chains, we expect to find a distribution of sizes. Having all topological edge-states at a fixed energy regardless of spin-chain size is an important step in their study and reproducibility.

In this Letter, we make a theoretical study of edge states on ferromagnetic (FM) spin chains on a BCS superconductor. Using the model presented in Ref.~\cite{Mier_topo}, we observe energy oscillations of the edge states as we vary the number of atoms in the magnetic chain. Similar results have been recently observed in experiments~\cite{Schneider2}. Here, we show that the oscillations about zero can be found for topological and trivial states of the chain. We trace back the origin of the periodicity to a 1D moir\'e pattern 
resulting from the discrete nature of the spin chain reflected in its lattice parameter $a_c$, and the substrate's Friedel oscillations 
governed by the substrate's Fermi wavelength, $\lambda_F$. As these two periods approach, the resulting edge-state energies rapidly change, producing a rich variety of patterns akin to the many different schemes of energy bands found in 2D moir\'e systems \cite{Moire_bans,lisi_observation_2021}.  The dependence of the edge-state energies on $a_c$ and $\lambda_F$ allows us to control their dispersion. In this way, we can engineer magnetic chains for which the topological edge states remain fixed at a finite energy regardless of the actual size of the spin chain beyond a minimal size.

To simulate the FM spin chain on an s-wave superconductor, we solve the Bogoliubov-de Gennes equations~\cite{Mariano} for a superconducting surface modelled as a 2D array of points, spaced by the lattice parameter, $a$, see  Refs.~\cite{Mier_topo,Mier_BiPd}. The superconductor is defined by the order parameter, $\Delta$, the Fermi vector of the material, $k_F$ and the metallic electron density, $N_0$. These parameters fix the BCS correlation length, $\xi$, that we use here. The effect of the magnetic impurities is modelled by the local Hamiltonian:
\begin{equation}
	\hat{H}_{impurity} = \sum_{j, \sigma}^N (K_j \hat{c}_{j\sigma}^\dagger\hat{c}_{j\sigma} + J_j\vec{S_j}\cdot\hat{\vec{s}}(j)),\;
	\label{Kondo}
\end{equation}
where $J$ is the magnetic exchange interaction of the impurities with the superconductor, $\Vec{S}_j$ is the spin of impurity $j$ that is assumed to be classical, $K_j$ is the corresponding non-magnetic scattering potential and $\hat{\vec{s}} (j)$ is the electron spin operator at the tight-binding orbital localized at $j$ corresponding to the creation and annihilation operators $\hat{c}_{j\sigma}^\dagger$ and $\hat{c}_{j\sigma}$. The exchange interaction of the magnetic impurities with the Cooper pairs in the superconductor results in the emergence of localized states inside the superconducting gap known as Yu-Shiba-Rusinov (YSR) states~\cite{yu_1965, Shiba_1968, Rusinov_1969}. By creating a chain of such impurities,  the localized states can hybridize creating YSR bands which can go into the topological phase~\cite{Pientka2013, Ojanen_2014, Andolina_2017}.
In case of FM chains, the system can be in the topological phase when the Rashba spin-orbit coupling, $\alpha_R$, is different from zero~\cite{Tewari2012}.

\begin{figure}[ht]
    \centering
    \includegraphics[width=0.5\textwidth]{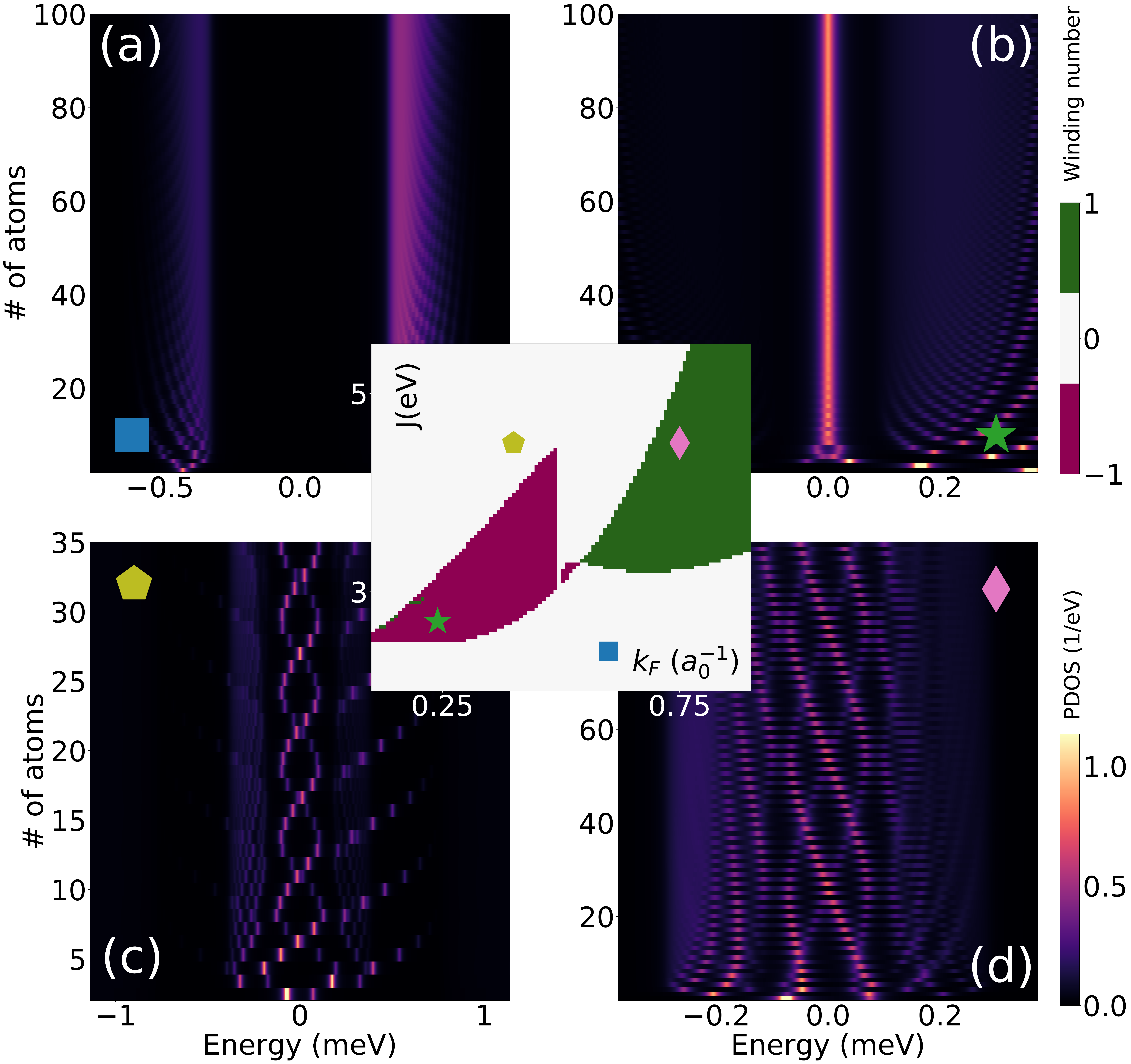}
	\caption{Topological phase diagram and evolution of edge states with number of atoms. Inset: Winding number calculated for infinite FM chains as a function of the magnetic coupling, $J$ and the Fermi vector, $k_F$. The green and magenta areas correspond to the topological phases $w=1$ and $w=-1$, respectively, and the white area is the trivial region. The shapes indicate pairs of parameters for which we investigate the evolution of edge states. (a)-(d) Projected density of states (PDOS) as a function of the number of magnetic atoms in the spin chain. The $J$ and $k_F$ parameters are indicated by the position of the corresponding shape. (a) $J=2.4$~eV, $k_F=0.6$ $a_0^{-1}$, blue square. (b) $J=2.7$eV, $k_F=0.24$ $a_0^{-1}$, green star. (c) $J=4.5$ eV, $k_F=0.4$ $a_0^{-1}$, yellow pentagon. (d) $J=4.5$ eV, $k_F=0.75$ $a_0^{-1}$, pink diamond. Other parameters are $\Delta = 0.75$ meV, $N_0=0.037$/eV, $\alpha_R = 3.0$ eV-\AA\, $K=5.5$ eV and $a_c=3.36$~\AA. In (d), the correlation length is the BCS one, $\xi=4852$~\AA.}
	\label{Fig_1}
\end{figure}

Figure~\ref{Fig_1} relates the energy behavior of edge states to the topological phase of the chain. In the inset of Fig.\ref{Fig_1}, we reproduce the phase space from Ref.~\cite{Mier_topo}. The winding number~\cite{Tewari2012}, $w$, of an infinite FM chain is plotted as a function of the magnetic coupling of the impurities, $J$, and the Fermi vector, $k_F$. In Ref.~\cite{Mier_topo} we showed that the winding number matches the behavior of the Pfaffian topological invariant and its use can be extended to the case of 2D-Rashba coupling. The winding number yields more information than the Pfaffian because it comes in three values for the present system~\cite{heimes_2015,Li_2018} with $w=0$ the topologically trivial phase and $w=\pm 1$ the non-trivial phases. The green and magenta areas on the phase diagram correspond to the topological phases of $w=1$ and $w=-1$, respectively. On this phase diagram, we select four pairs of parameters and study the evolution of the edge states as we change the length of the chain.

On Fig.\ref{Fig_1} (a)-(d) we plot the projected density of states (PDOS) on the edge atom of the chain  as a function of energy ($x$-axis) and chain length ($y$-axis). Figure \ref{Fig_1} (a) corresponds to a non-topological case  (square in the phase diagram) where the in-gap states are still far from zero energy and delocalized along the chain. As we increase the number of atoms, the edge states remain at fixed energy. On Fig.\ref{Fig_1} (b) we plot the in-gap states for a case in the topological region ($w=1$), here the system has undergone a topological phase transition (TPT) and a MBS appears at the edge of the chain for chains as short as five atoms.  On Fig.\ref{Fig_1} (c) the system has undergone through a new TPT, and it is again on the trivial phase. At the lower energies, states oscillate around zero as the number of atoms in the chain increases with a periodicity of \cris{$\sim 5$} atoms. These states are also localized at the edges, however the winding number shows they are topologically trivial. Finally, Fig.\ref{Fig_1} (d) depicts the evolution of edge states  in the $w=-1$ topological phase. Here, the oscillations have a higher periodicity of \cris{$\sim 60$} atoms as well as two different branches that can be separated into even and odd number of atoms. These results show that the presence or absence of oscillating edge states around zero cannot be linked to the topology of the spin-chain phase.
 
Friedel oscillations arise from the presence of impurities on different substrates with extended electronic states such as on superconductors~\cite{Rusinov_1969,Bena_Friedel,Menard_2015}. The oscillation follows the relation $\cos \, (2 k_F r +\delta)$, where $r$ is the electronic coordinate, $k_F$ the Fermi wave vector and $\delta$ a phase shift~\cite{Rusinov_1969}. However, if we try to apply this to understand the periodicity observed {in Fig.}~\ref{Fig_1}, the resulting oscillations should have a much shorter period than the ones we observe for the chains. Another surprising fact is that small changes of the inter-impurity distance or of $k_F$ can lead to very large non-monotonic variations of the period of the oscillations. To understand this behavior, we solve a simpler case.

\begin{figure}[ht]
    \centering
    \includegraphics[width=0.5\textwidth]{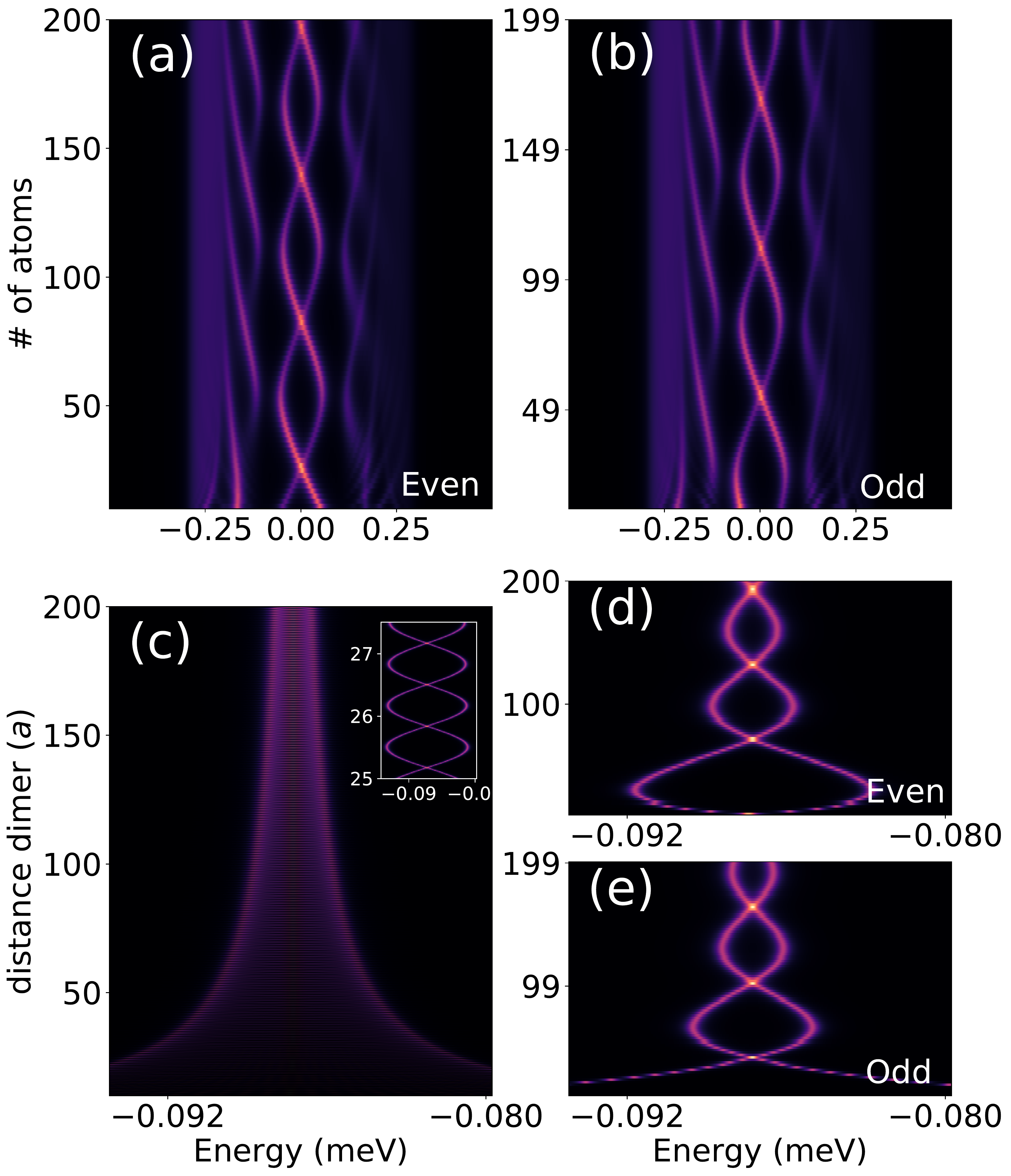}
	\caption{Projected density of states (PDOS) on the edge atom of a spin chain as a function of the number of atoms in the chain for (a) even and (b) odd number of atoms respectively. When even and odd numbers of atoms are plotted in the same graph, we recover Fig.~\ref{Fig_1} (d) ({$k_F=0.75$}~$a_0^{-1}$). (c) Dimer PDOS as the distance between the two atoms is varied in a continuous way. Inset: Zoom-in area between distances {25-27.5} $a_c$. The PDOS from (c) is plotted at selected distance in (d) and (e) corresponding to integer multiples of $a_c$, even and odd integers, respectively. This clearly shows that the observed oscillations are a consequence of the discrete sampling of the continuous PDOS in (c) and they match the periodicity of the oscillations in (a) and (b), revealing the moir\'e nature of the latter (the parameters are the same as in  Fig.~\ref{Fig_1} (d) and the above (a) and (b)).}
	\label{Fig_2}
\end{figure}

 We solve the Bogoliubov-de Gennes equations for a dimer of local magnetic impurities on a homogeneous s-wave superconductor. We create a two-site system with two magnetic impurities, and we vary continuously the inter-impurity distance, $r$. Figure~\ref{Fig_2} shows the comparison between chain and dimer for the chain of Fig.~\ref{Fig_1} (d) at a fixed lattice parameter $a_c=3.36$~\AA. {Figure}~\ref{Fig_2} (a) shows chains with an even number of atoms  and (b) an odd number. In contrast to the chain calculation, where the distance between sites is fixed by multiples of the lattice parameter of the superconductor, the calculation's distance between atoms in the dimer, $r$, is varied continuously from 10 to 200 lattice parameters, Fig.~\ref{Fig_2} (d), (e) and (f). The inset {in Fig.}~\ref{Fig_2} (d) shows a zoomed-in area for dimer distances between $25\, a_c$ and {$27.5\, a_c$}, where the spin-chain lattice parameter equals to lattice parameter of the superconductor, $a_c=a$. We find that the period of the oscillation is $0.66 a_c = 2.2$~\AA, matching the Friedel period $\lambda_F/2$ for $k_F=0.75 a_0^{-1}$ as corresponds to the oscillations of superconductor-mediated interactions. The amplitude of the oscillations decays as $1/r^2$, in excellent agreement with the decay of superconductor-mediated interactions between impurities on a 2D-Rashba s-wave superconductor~\cite{Malshukov}. 

To retrieve the period found on the spin chains, we need to probe the PDOS of the dimer at selected distances corresponding to an integer number of lattice parameters $a$. This is what we plot in Fig.\ref{Fig_2} {(d)} and {(e)}. As we can observe, the resulting oscillation has the exact same period as the one observed in Fig.~\ref{Fig_2} {(a)} and {(b)} {(a period of 60 atoms)} with the same even- and odd-site behavior. We have checked that we can understand the behavior of all oscillations in the phase space of Fig.~\ref{Fig_1} in the same way.

The observed oscillations can be rationalized by the presence of two periodical structures: one is the Friedel-like oscillation, with a wave vector of $k_1= 2 k_F$, and the other one is the discrete spatial sampling by the spin chain, controlled by the inter-impurity distance, $a_c$, giving a wave vector $k_2=2\pi/a_c$. The resulting periodic structure is a moir\'e pattern and the emerging periodicity  depends on the two parameters $k_F$ and $a_c$. As corresponds to moir\'e patterns, small changes in these two parameters can lead to large differences in the observed periods. In 1D, the moir\'e pattern coincides with the beat phenomenon of acoustics. Hence, the resulting periodic structure can  be approximated by the product of two harmonic functions with different frequencies: one with longer period and frequency $\frac{k_1-k_2}{2}$ that results in the envelope oscillation  of the signal encompassing tens of atoms in the above examples, and another with high frequency $\frac{k_1+k_2}{2}$ that is responsible for the even-odd effect, Figs.~\ref{Fig_2} {(a)} and {(b)}. Indeed, the second frequency is too high to be visible in the discrete lattice of the chain and thus, results in an alternating signal when we add atoms to the chain.
\\

\begin{figure}[ht]
    \centering
    \includegraphics[width=0.45\textwidth]{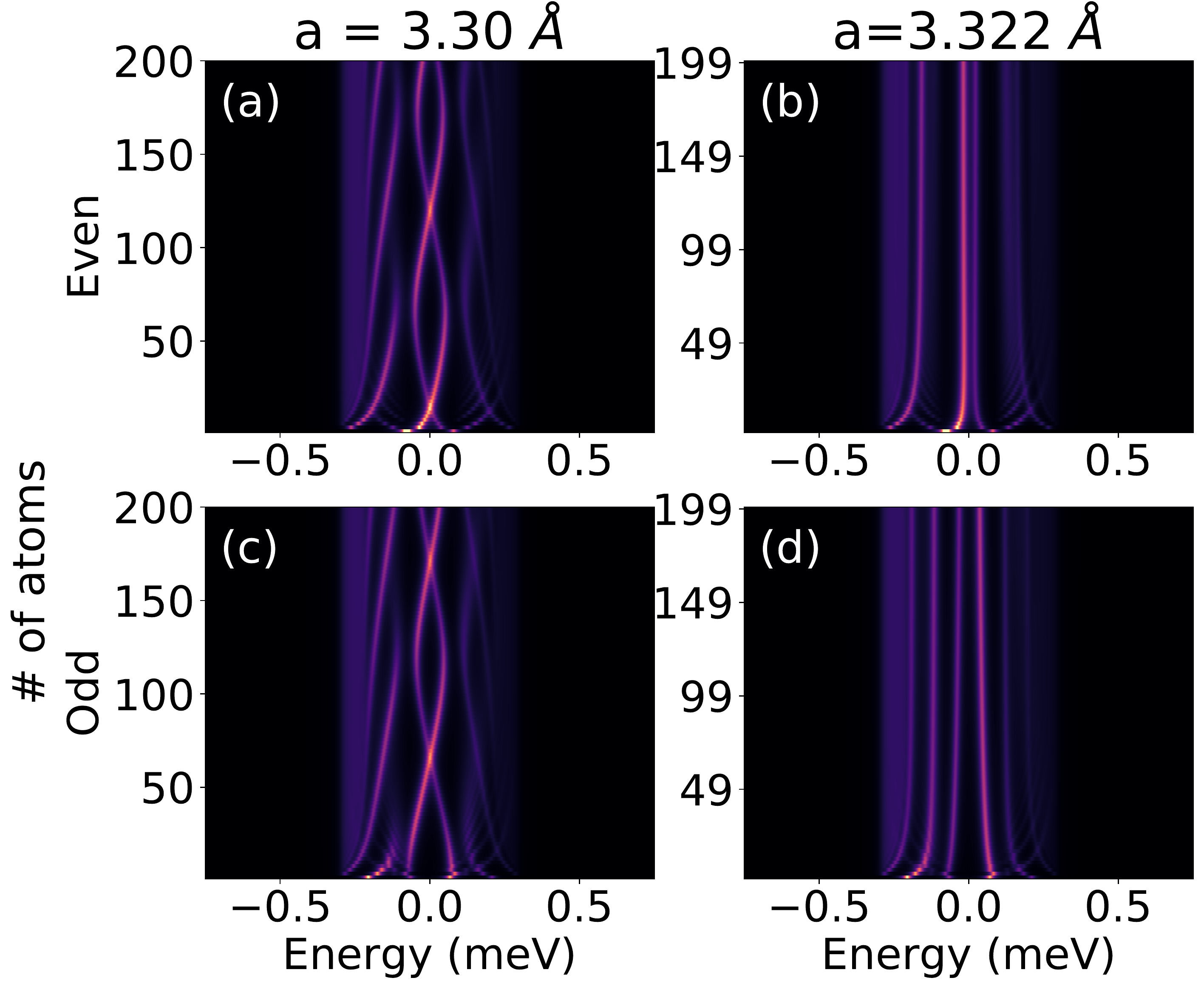}
	\caption{PDOS on the edge atom of spin chains for \textit{slightly} different inter-impurity distances, $a_c$. {In the right panels} $a_c=3.30$~\AA, the period of the resulting oscillations is {$\sim 107$} atoms, shown for even (a) and odd (c) number of atoms. In (b) and (d) $a_c=3.322$~\AA,  we do not observe any evolution of the edge states as we vary the number of atoms. These plots can be compared with Fig.~\ref{Fig_2} (a) and (b) $a_c=3.36$~\AA, where we find the period of the oscillations to be {60} atoms. The parameters other than $a_c$ are the same as in Fig.~\ref{Fig_1} (d).  
	}
	\label{Fig_3}
\end{figure}

The high dependence of the resulting pattern on $k_F$ and $a$ also allows for tuning the period of the oscillations.  We plot two different types of chain in Fig.~\ref{Fig_3}, using \textit{slightly} different lattice parameters. For $a_c=3.30$~\AA, we find again oscillating edge states that can be split into lengths of even and odd number of atoms. The period of the oscillations is found to be {$\sim 105$} atoms. For chains with $a_c=3.322$-\AA, we find two sets of in-gap states that can be again divided into even and odd, but the periods of these two oscillations are very large. If we increase roughly by 1\% the lattice parameter corresponding to the previous distance-between-impurities of $a_c=3.36$~\AA~(same chains as in Fig.~\ref{Fig_2} (a) and (b)) the  oscillations are found to have a much smaller period, $\sim 60$ atoms. As a result of the underlying moir\'e pattern, small changes in $a_c$ lead to drastically different periods.  

\begin{figure}[ht]
    \centering
    \includegraphics[width=0.45\textwidth]{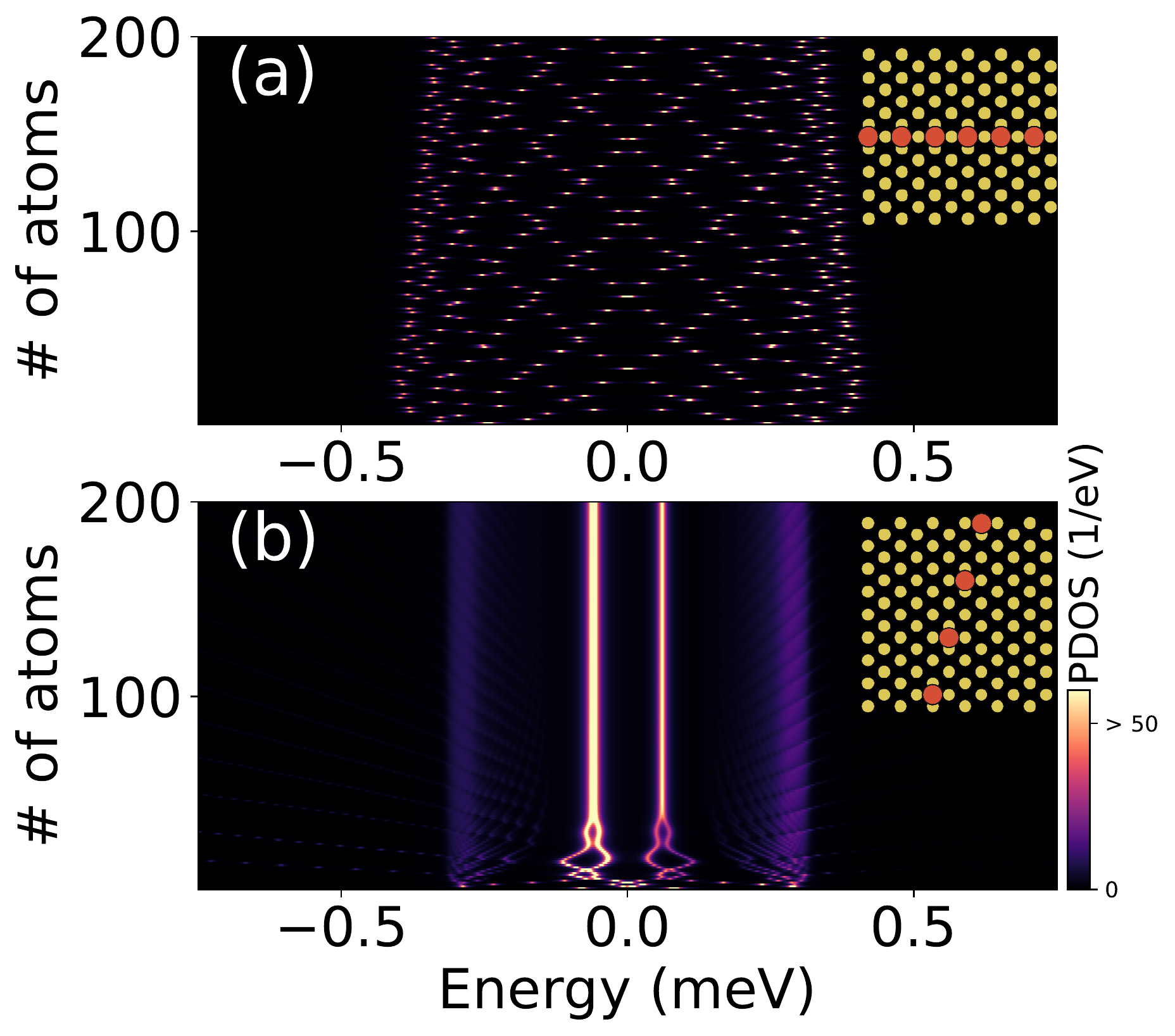}
	\caption{PDOS obtained on the edge atom of Mn chains of increasing length on Nb (110). (a) The atomic chains are oriented along the $[1\Bar{1}0]$ direction, resulting in a distance between impurities of $a_c=4.67$ \AA. (b) The chains are oriented along $[\Bar{\frac{1}{2}}, \frac{1}{2},\frac{5}{2}]$ direction, where the distance between impurities is $a_c=8.56$ \AA. The parameters of the calculation are $k_F=0.1886$ $a_0^{-1}$, $J=0.125$~eV, $\Delta = 1.5$~meV, $N_0=1.0$~eV$^{-1}$, $\alpha_R = 1.0$~eV-\AA, $K=0.063$~eV, and $a=3.294$~\AA. The BCS correlation length is $\xi=576$~\AA. The insets show schemes of the crystal surface and the chain orientations.}
	\label{Fig_4}
\end{figure}

A practical application of our calculations is  to reproduce the experimental results by Schneider et al. \cite{Schneider2}. These authors studied the in-gap states on Mn chains built along the $[1\Bar{1}0]$   direction of superconducting Nb (110) surface. They observed energy oscillations about zero as they increase the number of atoms in the chain with an even-odd behavior similar to the behavior of Fig.~\ref{Fig_2} (b) and (c). On Fig.~\ref{Fig_4} (a) we plot the calculated PDOS on the edge atom obtained using parameters extracted from Ref.~\cite{Schneider2}. Our results are in overall agreement with the experiment, in particular reproducing the even-odd behavior and the finite energy-range of the in-gap states. However,  if we compute the in-gap electronic structure when the spin chains are oriented along the $[-1/2, 1/2, 5/2]$ 
direction, we obtain the results of Fig.~\ref{Fig_4} (b). As we can observe, in this case the edge states do not evolve for chains longer than $\sim 20$ atoms, and they remain at a fixed finite energy. This behavior is a consequence of the moir\'e pattern when $k_1-k_2=0$. Here, we have shown that we can tailor the dispersion of edge states with spin-chain length by choosing the growth direction of the spin chain on the surface. 
\\

In summary, solving Bogoliubov-de Gennes equations for a local model of classical magnetic impurities on an s-wave superconductor predicts oscillations in the energy of edge states as a function of the length of the impurity chain. The origin of the period of the oscillation is a moiré pattern resulting from the discrete sampling produced by the spin chain on the superconductor-mediated impurity-impurity interaction. The sampling produces a natural spatial frequency of $2\pi/a_c$ where $a_c$ is the inter-impurity distance, and the substrate-induced interaction has a Friedel-like oscillation with the substrate's Fermi vector, $k_F$. The origin of the edge-state oscillation is thus independent of its topological character. We have found that by modifying the spin-chain geometry, it is possible to tune the moir\'e pattern to display edge states  at fixed finite energy. We predict that these non-dispersive states can be realized on Mn chains on Nb (110) crystal by assembling a spin chain along the $[-1/2, 1/2, 5/2]$  direction of the Nb (110) surface. It is then possible to create edge states with an important topological character~\cite{Schneider2} at fixed finite energies that are independent of the distribution of lengths in the creation of spin chains. The idea of having topological states at finite energy opens the door for a wider range of experiments involving MBS. 

Financial support from the Spanish MICINN (projects RTI2018-097895-B-C44
and Excelencia EUR2020-112116) and Eusko Jaurlaritza (project
PIBA\_2020\_1\_0017) is gratefully acknowledged.

\bibliography{refer_oscil}
\end{document}